\documentclass[prb,twocolumn]{revtex4}
\usepackage{amsmath,amssymb,graphicx}
\begin{document}
\title{A Hybrid Density Functional Study of Oligothiophene/ZnO Interface for Photovoltaics}
\author{Na Sai,$^1$ Kevin Leung,$^2$ James R. Chelikowsky$^3$}
\affiliation{$^1$Center for Nano and Molecular Science and Technology, University of Texas, Austin, Texas, 78712, USA\\
$^2$Surface and Interface Sciences Department, MS1415, Sandia National Laboratory, Albuquerque, New Mexico 87185, USA \\
$^3$Center for Computational Materials, Institute of Computational Engineering and Sciences, 
Departments of Physics and Chemical Engineering,  University of Texas, Austin, Texas, 78712 USA}
\date{\today}
\begin{abstract}
Organic/inorganic donor-acceptor interfaces are gaining growing attention in organic photovoltaic applications as each component of the interface offers unique attributes. Here we use hybrid-density functional theory to examine the electronic structure of sexithiophene/ZnO interfaces. We find that interfacial molecular orientations strongly influence the adsorption energy, the energy level alignment, and the open circuit voltage. We attribute the orientation dependence to the varied strength of electronic coupling between the molecule and the substrate. Our study suggests that  photovoltaic performance can be optimized by controlling the interfacial design of molecular orientations.  
\end{abstract}
\maketitle 

Organic photovoltaics (OPV)  differ from conventional solar cells in that their fundamental mechanism is dominated by interfacial rather than bulk processes.\cite{Brabec,Gregg}  The primary energy conversion process in OPVs involves excitation of tightly bound electron-hole pairs (excitons) and charge carrier generation, {\it i.e.}, a separation of charges across a heterointerface formed by an electron donor and acceptor (e.g., a conjugated polymer and a functionalized fullerene~\cite{Yu}). Because of the large exciton binding energies, the OPV system must provide a suitable energetic driving force for the excitons to dissociate, making charge separation a key process that limits OPV efficiency. 
Owing to their low cost and scalability, photovoltaics based on polymer and small molecule organics have received increasing attention in recent years.~\cite{Coakley, Lloyd} However, basic understanding of the fundamental processes at the interfaces, such as charge separation and charge transfer, continues to lag behind that for inorganic PVs. 

An attractive alternative to all-organic solar cells is a hybrid solar cell that mixes an  organic/polymeric system with an inorganic semiconductor/metal oxide such as CdSe,~\cite{Huynh} TiO$_2$,~\cite{Coakley03} or ZnO~\cite{Law} as the electron acceptor. These inorganics are suitable for photovoltaic applications owing to their relatively high electron affinity and mobility. The flexibility to pattern them into nanocrystalline structures makes them desirable for highly ordered nanostructured photovoltaic cells.~\cite{Briseno}
There have been many reports of hybrids combining poly(3hexylthiophene)(P3HT) and ZnO.~\cite{Beek06} 
Their efficiencies, however, are much lower than those of fullerene based organic blends.  Identifying the factors limiting the efficiencies of organic/inorganic solar cells remains a challenge.

Owing to their highly ordered and controllable crystalline structures, oligomers such as sexithiophene (6T) offer an ideal system to study the electronic structure at organic/inorganic interfaces.~\cite{Duhm, Garcia, Loi, Ivanco, Blumstengel08} Blumstengel {\it et al.} have recently reported atomic force and ultraviolet photoelectron spectroscopy (UPS) of 6T on ZnO surface showing that the heterostructure holds promise for photovoltaic applications.~\cite{Blumstengel08}  Depending on the surface metallicity, directions, and coverage, the 6T molecules can take either a flat ``lying-down'' or an upright ``end-on'' orientation.~\cite{Loi, Ivanco, Blumstengel08} How the molecular orientations affect the interfacial electronic structures is not yet completely understood.  

Here we carry out a first principles study of the sexithiophene and ZnO interface using hybrid functional density-functional theory (DFT). 
Our calculations address the question of orientation dependence and show that both the upright and flat-lying orientations give rise to type-II interfacial band alignment that is necessary for charge separation. Their interfacial energy offsets are, however, very different. Controlling the orientations of organic oligomers on inorganic substrates offers an opportunity for molecular engineering of the interface so as to achieve good photovoltaic performance. 

We carry out calculations using a plane-wave pseudopotential approach to DFT implemented in the {\tt VASP} code~\cite{VASP}  with PAW potentials.~\cite{PAW} To study electronic properties we employ the PBE0 hybrid-functional~\cite{Perdew96, Adamo} in which a fraction of the nonlocal exact exchange energy is included in the exchange-correlation potentials. Although not completely free of the self-interaction error,~\cite{Sanchez} hybrid functionals have yielded much improved band gaps for semiconductors including ZnO.~\cite{Paier, Oba, Labat} Computational methods that minimize the error in bandgap calculations are essential for accurate theoretical predictions of interfacial energy alignment involving organic or inorganic systems.~\cite{Neaton, Grossman,Alkauskas} We use $6\times6\times6$, $\Gamma$-point, and $1\times2\times 1$ for $k$-point sampling of bulk ZnO and the two supercells, respectively, and a plane-wave energy cutoff of 400 eV. 

Using the PBE generalized gradient approximation,~\cite{Perdew} we obtain lattice constants of $a =3.29$\AA, $c = 5.30$\AA~  (exp. $a=3.25$\AA, $c = 5.21$\AA\cite{Madelung}) that agree with previous studies~\cite{Meyer03}and a band gap of 0.7 eV for bulk ZnO. The PBE0 hybrid functional gives rise to band gaps of 3.08 eV and 3.16 eV for the bulk and the ({10$\bar{1}$0}) surface, respectively, in much better agreement with the experimental 3.37 eV.~\cite{Madelung}    For the 6T molecule, we have obtained an energy gap of 1.57 eV in PBE and 2.7 eV in PBE0 calculations (at the PBE geometry), the latter compares well with the experimental value~\cite{Izumi} of 2.8 eV and an earlier B3LYP hybrid functional calculation.~\cite{Zade}

\begin{figure}
\includegraphics*[width=8.5cm]{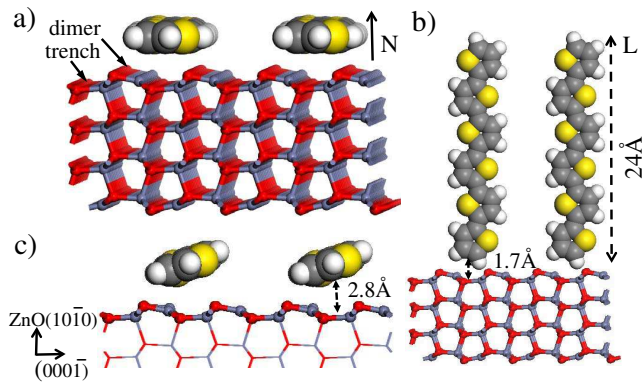}
\caption{(color online) Atomic structure of the 6T/ZnO (10$\bar{1}$0)  interface where the 6T molecule is in (a) a flat-lying ($N$) and (b) an upright standing ($L$) orientation. (c) The interfacial structure after the relaxation for the $N$ interface. The dashed arrows mark the average distance of the molecule above the ZnO trench rows. }
\label{interface}
\end{figure} 
Although the growth of 6T on both nonpolar and polar ZnO surfaces has been demonstrated,~\cite{Blumstengel08} the structure of polar ZnO surfaces, and in particular the charge passivation and surface reconstruction mechanism, are still under debate.~\cite{Meyer04,Ding} Here we consider the uncharged nonpolar (10$\bar{1}$0) surface that has the lowest cleavage energy and is easy to work with experimentally.~\cite{Diebold} We have considered a ZnO (10$\bar{1}$0) surface slab  including four ZnO double layers  (see Fig.~\ref{interface}). Each surface unit cell contains a ZnO dimer row separated by a trench row along the $[000\bar{1}]$ direction. Surface relaxation has led to a strong tilt of the ZnO dimer as a result of the anions moving closer to the bulk, similar to what has been reported earlier.~\cite{Meyer03}  The upright standing ($L$) and flat-lying ($N$) interfaces are constructed by aligning the long (backbone) and normal axis of the molecule, denoted by $L$ and $N$,  to the surface normal, as illustrated in Fig.~\ref{interface} (a) and (b).  Owing to the incommensurate surface lattice parameters for 6T crystal ($a$ = 6.0~\AA~ and $b$ = 7.9~\AA)~\cite{Yassar} and ZnO (10$ \bar{1}$0) surface ($a$ = 3.288~\AA~ and $b$ = 5.303~\AA) along with a periodic boundary conditions, modeling the interface with a 6T and matching ZnO slab  would introduce an unrealistically large strain. Instead we have adopted a supercell that contains $2\times2$ and $10\times2$ ZnO surface unit cells for the $L$ and $N$ orientation, respectively. This neglects the intermolecular interactions.
Each supercell contains one 6T molecule initially placed above the trench row. Recent calculations of Dag \& Wang within the local density approximation (LDA) have shown this position to be energetically favorable for a  flat-lying P3HT molecule on ZnO (10$\bar{1}$0) surface.~\cite{Dag} We include a vacuum of more than 8 \AA~ along the interface direction to ensure isolation of the interface.

Starting with a fully relaxed ZnO (10$\bar{1}$0) surface, we initially position the 6T molecule at a height of 2-3\AA~ (depending on the orientation) above the trench and relaxed the atomic positions of the molecule and the first two ZnO surface double layers while fixing position of the bottom ZnO layers. Upon the relaxation, the 6T molecule undergoes about $20^\circ$ tilting relative to the surface in the $N$ orientation due presumably to interaction between the molecule $\pi$ face and the tilted Zn-O dimers, as shown in Fig.~\ref{interface} (c), while remaining perpendicular to the surface in the $L$ orientation. The final position of the molecule has an average adsorption distance of 2.8 and 1.7~\AA~ for the $N$ and $L$ orientations, respectively, above the surface. The former value is comparable to that of P3HT/ZnO.~\cite{Dag}  Because of the computational cost of the hybrid functional calculation (which is up to 30-50 times more expensive than with the standard functionals) and a large supercell size, we have kept the interfacial geometries at the PBE level which has been shown to yield reasonable surface geometries.~\cite{Meyer03}
Van der Waals (vdW) interactions which are absent in the PBE functional may influence the molecular tilting and adsorption distance.~\cite{Tkatchenko10}  Tests indicate, however, that removing the tilting and reducing the adsorption distance by up to 0.3 \AA~ produce essentially the same electronic structure in our results.

The adsorption energy of 6T on the ZnO interface is calculated from the difference between the total energy of the interface and that of the fully relaxed ZnO surface and 6T molecule. Within the PBE functional, we have obtained an adsorption energy of $\sim 1$eV for the $N$ and 0.1 eV for the $L$ orientation. 
(The PBE0 functional gives 2 eV and 0.4 eV, respectively, at the PBE geometry.) The former value seems much smaller than the binding energy of 2.8-2.9 eV in ref.~\onlinecite{Dag}. 
Owing to the lack of vdW interactions, neither PBE (PBE0) nor LDA functional gives an accurate account of the adsorption binding energy for physisorbed molecular interfaces.~\cite{Tkatchenko10} 
An explicit inclusion of the van der Waals (vdW)  interaction is, however, computationally demanding and well tested parameters are lacking for systems involving metal oxides.~\cite{Dion, Tkatchenko}   Here we focus on the relative stabilities of different molecular orientations. Compared to the upright standing orientation, the flat-lying one is much more energetically stable owing to a stronger orbital overlap between the molecular $\pi$ wavefunctions and the substrate surface. This relative stability applies to the situation in which the molecule--substrate interaction is dominant (e.g., in a low coverage interface), but but the upright standing orientation would be stabilized if intermolecular interactions are taken into account.

\begin{figure}
\includegraphics[width=3.5in]{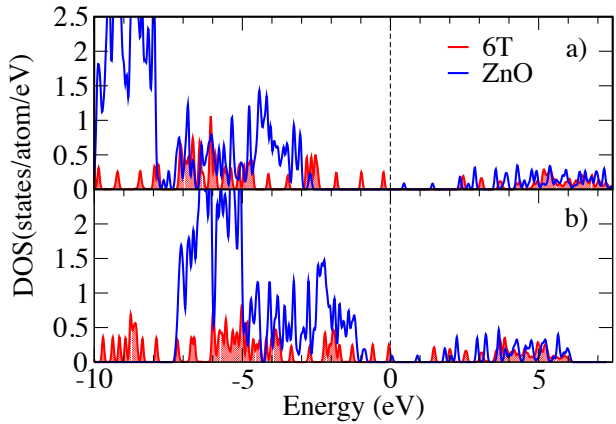}
\caption{(color online) Projected density of states on the ZnO (blue solid lines) and 6T orbitals (red solid lines shaded from below) of the upright standing interface obtained from the (a) PBE0 hybrid and (b) PBE functional. The dashed line denotes the Fermi energy which is at the same level of the 6T HOMO.}
\label{PDOS_PBE0}
\end{figure} 
We apply the PBE0 hybrid functional to study the interfacial electronic structures. In Fig.~\ref{PDOS_PBE0}(a) we plot the projected density of states (PDOS) of the upright standing interface. The Zn $3d$ band lies 6 eV below, and is well separated from the top of the valence bands which are composed of hybridized Zn $3d$ and O $2p$ states. The bottom of the conduction band (Zn $4s$) lies 3 eV above the valence band maximum (VBM). The 6T states are discrete, with the highest occupied molecular orbital (HOMO) level lying in the ZnO gap and the lowest unoccupied molecular orbital (LUMO) level about 2.5 eV above the Fermi level overlapping with the ZnO conduction bands. Fig.~\ref{PDOS_PBE0}(b) plots the PDOS calculated from the PBE functional. The Zn $3d$ states are much shallower in energy, this pushes up the O $2p$ states and leads to a much smaller gap in ZnO compared to the case of PBE0. The ZnO conduction band minimum (CBM) also lies much lower in energy than in the PBE0 case and overlaps with the 6T HOMO at the Fermi level. The band offset predicted using the PBE functional is unphysical.  Henceforth we focus on the PBE0 results.

\begin{figure}
\includegraphics[width=3.5in]{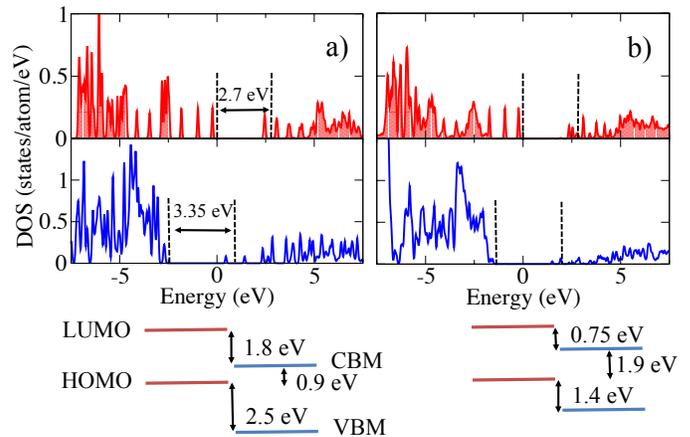}
\caption{(color online) Projected density of states for the (a) $L$ oriented and (b) $N$ oriented interface. The vertical dashed line marks the 6T HOMO and LUMO and the ZnO VBM and CBM. The LUMO and CBM are obtained by adding the corresponding optical gap for 6T (2.7 eV) and ZnO (3.35 eV) to the occupied levels. Below the panels are calculated energy offsets for
the 6T/ZnO interfaces.}
\label{levels}
\end{figure}

 A suitable energy offset at the donor and acceptor interface is a key aspect for charge separation and charge transfer in OPVs. To calculate the level alignment at the 6T/ZnO interface, we focus on the PDOS near the HOMO and LUMO. For the upright standing ($L$) interface as shown in Fig.~\ref{levels} (a), the 6T HOMO and LUMO levels are both higher in energy than the ZnO VBM and CBM respectively, indicating that the interfacial alignment is a staggered type-II interface ideal for organic photovoltaics. By comparing the energy positions, we find that the $\Delta E_{\rm HOMO/VBM}$ and $\Delta E_{\rm LUMO /CBM}$ offsets are of 2.5 eV and 2 eV. 
From the UPS measurement,~\cite{Blumstengel08}  Blumstengel {\it et al.} have reported a $\Delta E_{\rm HOMO/VBM}$ of 2.15 eV and a $\Delta E_{\rm LUMO /CBM}$  of 1.5 eV for 6T/ZnO (000$\overline{1}$).
The latter offset was obtained by adding the optical energy gap, which take into account of the exciton binding energy of 6T and ZnO, to the occupied levels. Using the same optical gaps,  we have obtained a $\Delta E_{\rm LUMO /CBM}$ offset of 1.8 eV. Despite the absence of intermolecular interactions, our data compares well with the UPS result and confirms that the experimental band offsets arise from upright standing molecules. The small differences maybe attributed to the surface dipole on ZnO(000$\overline{1}$) which affects the band offset. 
Compared to the exciton binding energy of 6T molecules ($\sim$ 0.4 eV), the LUMO/CBM offset is significantly higher, indicating a sufficient driving force for exciton separation across this interface. 

The flat-lying ($N$) interface, as shown in Fig.~\ref{levels}(b), also offers a type-II interface. However, the HOMO/VBM and LUMO/CBM offsets, 1.4 eV and 0.75 eV (0.84 eV from PDOS), respectively, are almost 50\% lower than those for the $L$ interface, albeit still sufficient for charge separation.  The percentage change of the offsets with respect to the molecular orientaion compares very well in magnitude with the recent experimentally measured values for sexiphenyl (6P)/ZnO interfaces.~\cite{Blumstengel10}  We further extract the open circuit voltage ($V_{\rm oc}$) that plays a signiciant role in photovoltaic efficiency. This is defined as the difference between the quasi-Fermi levels of the separated holes and electrons and can be approximated by the offset between the HOMO of the donor and the CBM of the acceptor. For the $N$ oriented interface, we find a $V_{\rm oc}$ of 1.9  eV which is significantly higher than 0.9 eV for the $L$ oriented interface.  Our results suggest a sensitive molecular orientation dependence of the energy offsets at hybrid interfaces. A control of the molecular orientations should provide an effective approach to increase the output potential and the OPV efficiency. 

\begin{figure}
\includegraphics*[width=8cm]{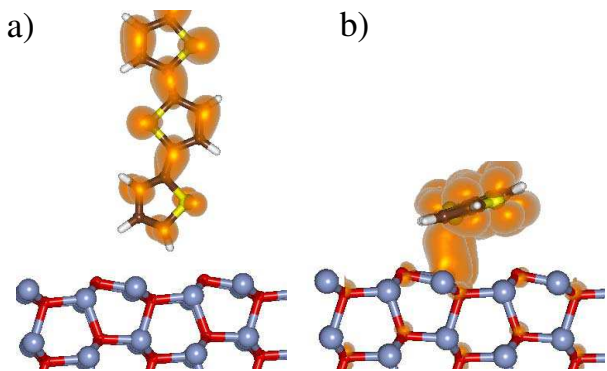}
\caption{(olor online) Isosurface of the electron density (the isovalue equals to the average of density) of the 6T LUMO on the (a) upright and (b) flat-lying interface. Panel (a) only shows half the number of thiophene rings in 6T.}
\label{LUMO}
\end{figure}
To understand the orientation dependence of the energy offsets, we carried out an analysis of the frontier orbitals of the 6T molecule at the upright ($L$) and flat-lying ($N$) interfaces. For both cases, the 6T HOMO is localized in the ZnO band gap while the LUMO overlaps with the ZnO conduction band in energy. This overlap has important implications in the electron transfer across the interface.  Fig.~\ref{LUMO}(a) shows the spatial distribution of the electron density of the LUMO at the upright interface.  The LUMO is a $\pi^*$ orbital and is almost completely localized on the molecule chain and displays little electronic coupling with the ZnO substrate despite the overlap in energy. 
In contrast, the LUMO orbital at the flat-lying interface, as shown in Fig.~\ref{LUMO}(b), delocalizes over both the molecule and the substrate  (especially on the molecular side closer to the surface because of the tilt of the thiophene rings against the surface), indicating a strong electronic coupling between the molecule and the substrate at the interface.  The fact that the electronic coupling is much stronger for the flat-lying case compared to the upright case is consistent with the higher adsorption energy we find for the $N$ orientation. We can also understand the orientation dependence of the interfacial energy offsets and value of $V_{\rm oc}$ qualitatively from the strength of the interfacial electronic coupling.  In case of the flat-lying interface, the molecular orbital levels have shifted significantly closer to the substrate levels as a result of the stronger coupling.  This significantly reduces the LUMO/CBM and HOMO/VBM offset between the two materials while raising the $V_{\rm oc}$.  

In summary, we have employed a hybrid density functional to study the electronic structure of sexithiophene/ZnO interfaces. The upright and flat-lying orientations both offer suitable type-II interface that are ideal for charge separation across the interface. However the magnitudes of the energy offsets at the interface are strongly dependent on the molecular orientations. The calculated offsets for the upright orientation agree well with the experimental values obtained from UPS,~\cite{Blumstengel08} and the relative change of the offsets with respect to the orientations also compares fairly well with the measured change for 6P/ZnO interfaces.~\cite{Blumstengel10} Our analysis suggests that the varied strength of the electronic coupling between the molecule and the substrate should account for the orientation dependence of the energy offsets.

We thank N. Koch, X.Y. Zhu, J.W.P. Hsu, and J. Moussa for useful comments and discussions. This work is supported as part of the program ``Understanding Charge Separation and Transfer at Interfaces in Energy Materials (EFRC:CST)'', an Energy Frontier Research Center funded by the U.S. Department of Energy Office of Basic Energy Sciences (DOE/BES) under Award No. DE-SC0001091. KL is also supported by the DOE under Contract DE-AC04-94AL85000. Sandia is a multiprogram laboratory operated by Sandia Corporation, a Lockheed Martin Company, for the U.S. DOE.  JRC acknowledges support from the U. S. DOE/BSE and Office of Advanced Scientific Computing Research under grant No. DE-SC0001878.  Computational resource has been provided by the Texas Advanced Computing Center.

\end{document}